\def\CII{\mbox{C\,{\sc ii}}} 
\def\CIV{\mbox{C\,{\sc iv}}} 
 
\def\OI{\mbox{O\,{\sc i}}} 
\def\SiII{\mbox{Si\,{\sc ii}}} 
\def\SiIV{\mbox{Si\,{\sc iv}}} 
\def\keV{\mbox{ke\kern-.1emV}} 

\documentclass{aa} 
\usepackage{graphicx} 
\begin{document} 


\title{Evidence for chemical evolution in the spectra of high 
redshift galaxies
\thanks{Based on observations obtained with 
the FORS Instruments at the ESO VLT, Paranal, Chile on the course of the 
observing proposals 65.O-0049(A), 66.A-0547(A), 68.A-0013(A), 68.A-0014(A)}} 

\author{D.\,Mehlert\inst{1} 
\and S.\,Noll\inst{1} 
\and I.\,Appenzeller\inst{1} 
\and R.\,P. Saglia\inst{2} 
\and R.\,Bender\inst{2} 
\and A.\,B\"ohm\inst{3} 
\and N.\,Drory\inst{2} 
\and K.\,Fricke\inst{3} 
\and A.\,Gabasch\inst{2} 
\and J.\,Heidt\inst{1} 
\and U.\,Hopp\inst{2} 
\and K.\,J\"ager\inst{3} 
\and C.\,M\"ollenhoff\inst{1} 
\and S.\,Seitz\inst{2} 
\and O.\,Stahl\inst{1} 
\and B.\,Ziegler\inst{3} 
} 

\offprints{D.\,Mehlert,\\ 
\email{dmehlert@lsw.uni-heidelberg.de}} 

\institute{Landessternwarte Heidelberg, K\"onigstuhl, 
  D-69117 Heidelberg, Germany 
\and Universit\"atssternwarte M\"unchen, Scheinerstra{\ss}e~1, 
  D-81679 M\"unchen, Germany 
\and Universit\"ats-Sternwarte G\"ottingen, Geismarlandstra{\ss}e~11, 
  D-37083 G\"ottingen, Germany
}

\date{Received June 14; accepted July 2, 2002} 

\authorrunning{Mehlert et al.} 

\abstract{Using a sample of 57 VLT FORS spectra in the redshift range
$1.37<z<3.40$ (selected mainly from the FORS Deep Field
survey) and a comparison sample with 36 IUE spectra of local 
($z \approx 0$) starburst galaxies we derive \CIV\ and \SiIV\
equivalent width values and estimate metallicities of starburst galaxies as a
function of redshift. Assuming that a calibration of 
the \CIV\ equivalent widths in terms of the metallicity based on the
local sample of starburst galaxies is applicable to
high-$z$ objects, we find a significant increase of the 
average metallicities from about $ 0.16\,Z_{\odot}$ at the cosmic
epoch corresponding to $z \approx 3.2$ to about $ 0.42\,Z_{\odot}$ 
at $z\approx 2.3$. A  significant further increase in metallicity 
during later epochs cannot be detected in our data. 
Compared to the local starburst galaxies our high-redshift objects tend to be overluminous for a 
fixed metallicity. Our observational results are in good agreement with published 
observational data by other authors and with theoretical predictions
of the cosmic chemical evolution.  
}

\maketitle 

\section{Introduction}
\label{Introduction}

\begin{figure*}
\includegraphics{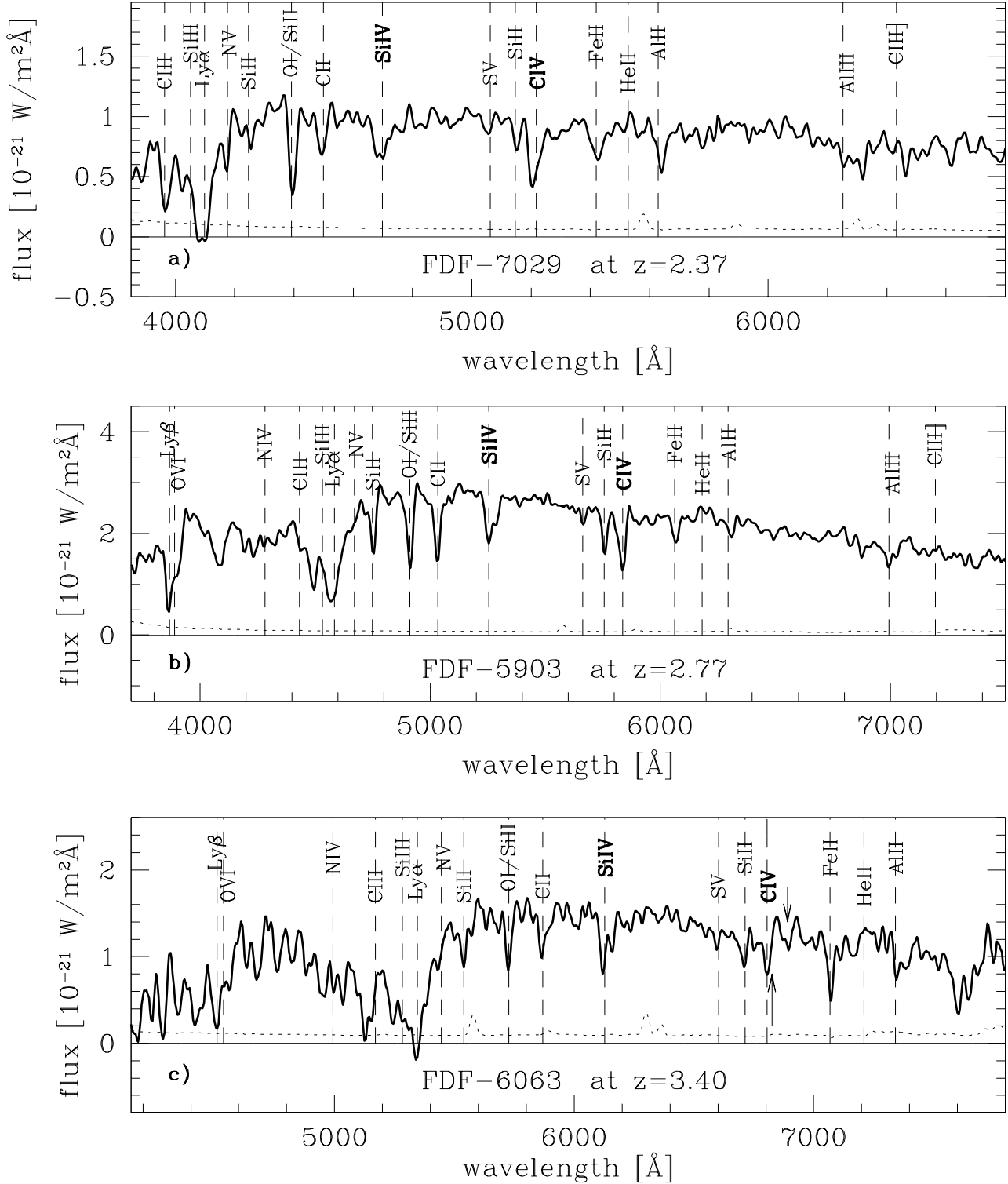} 
\caption{ 
Examples of the low-resolution galaxy spectra obtained with FORS.
The redshift of the objects is increasing from panel {\bf a)} to
{\bf c)}. The dotted line indicates the noise level which  (due to the 
night sky spectrum and the wavelength dependent instrumental efficiency)
varies with wavelength. The S/N of these spectra is 
about $\le$\,16, 43 and 12 for object 7029, 5903 and 6063, respectively. 
The object designations correspond to the catalog of Heidt et al. (2002).
For orientation the expected 
positions of selected spectral lines are indicated by 
vertical dashed lines. The two 
apparent emission features (marked by arrows) longward of 
the \CIV\ absorption feature are artifacts and resulted from the 
increased noise at this wavelengths due to telluric OH band edges and 
atmospheric absorption. For the same reason Fig. 1c becomes
unreliable beyond 7000 \AA .
\label{fig:exa_spectra} 
}
\end{figure*} 
Among the unsolved questions of cosmology is the beginning and
evolution of the star formation process at early cosmic epochs. 
From the presence of heavy nuclei in high-z quasars and galaxies it
is clear that star formation started rather early (see e.g.
Hamann and Ferland 1999, Dietrich et al. 1999). Moreover,
galaxy counts and emission line studies of high-z galaxies indicate that the 
star formation rate (SFR) declined rapidly since at least the epoch
corresponding to redshifts of $z \approx 2$ (see e.g. Madau et
al. 1996, Madau 2000). On the other hand intrinsic interstellar
extinction (which may affect strongly  the rest-frame UV of distant
galaxies) and uncertainties concerning the relation between line
emission and the star formation introduce considerable
uncertainties into the derivation of the global star formation
rate at early cosmic epochs from visual galaxy counts and
emission line studies, while investigations using IR and mm-wave data
are hampered by small samples. 

An alternative approach to 
investigate the early star formation history of the universe is the
evaluation of the chemical enrichment history of the universe, as 
star formation and the rapid evolution of massive stars results in a 
production of heavy nuclei more or less proportional to the SFR.  
Steidel et al. (1996a, 1996b) and  Lowenthal et al. (1997) have demonstrated 
that galaxies with redshifts up to about $z \approx 5$ can 
be observed with optical photometry and spectroscopy during periods 
of high star formation activity (resulting in high rest-frame UV emission
which is redshifted into the optical wavelength range in the observers
frame).  
Basic properties of these objects, such as number densities, luminosities, 
colors, sizes, morphologies, star formation rates, overall 
chemical abundances, dynamics and clustering have been investigated 
in various recent papers (cf. e.g. Steidel et al. 1996b, Yee et al. 1996, 
Lowenthal et al. 1997, Pettini et al. 2000, and Leitherer et
al. 2001). Steidel et al. (1996a, 1996b) already noted that the high-z 
galaxies show, on average, relatively weak metallic absorption lines,
and they ascribe this finding tentatively to
a lower metal content (a suggestion which has later
been reiterated in several subsequent papers). On the other hand 
most of the published spectra of high-z galaxies are not of sufficient S/N
to provide qualitative information on the metal content. Therefore,
we obtained new high S/N spectra of galaxies with $z \le 3.5$ with the
aim of  studying the chemical evolution of starburst galaxies at high
redshifts.

In the present paper we describe results on the \CIV\ absorption line
strength and their interpretation in terms of chemical evolution 
with cosmic age at redshifts $0<z<3.5$. Most of the new spectra were
obtained 
with the FORS instruments at the ESO VLT in the course of a 
photometric and spectroscopic study
of distant galaxies in the FORS Deep Field (FDF) 
(Appenzeller et al. 2000, Heidt et al. 2001, Bender et al. 2001). Due to its 
combination of depth and a (compared to the HDFs) relatively large area 
the FDF is particularly well suited for statistical studies of
high-redshift galaxies. In the present investigation we restrict
ourselves to redshifts $z \leq 3.5$ since at higher redshifts the position
of the redshifted \CIV\ resonance lines tends to coincide with strong
OH night sky lines.
Hence, for an accurate sky subtraction a higher spectral resolution or longer
exposure times than we could achieve so far would be needed. In Sect.~2 
we describe the 
sample selection and the observations. In Sect.~3 we present and discuss the 
measurement of the \CIV\ (and \SiIV ) equivalent widths, from which we estimate
the metallicities of the investigated objects in Sect.~4. In Sect.~5 we 
compare our results with data available in the literature, in Sect.~6 we draw 
our conclusions.

\section{Sample selection and observations}
\label{Sample}

The data set used in this study combines spectra of high-redshift galaxies
observed with the ESO VLT and spectra of local starburst 
galaxies taken from the IUE archive. 51 of the high-redshift 
spectra were selected (according to the criteria listed below) from 
about 300 low-resolution spectra observed mainly during the 
spectroscopic observing runs of the FDF program 
(in 3 nights in Sept. and Oct. 2000 and 3.5 nights in Oct. and Dec. 2001). 
For these
observations we used FORS1\&2 at the VLT in MOS and MXU mode with a 
slit width of 1$^{\prime\prime}$ and the 150I grism. (For instrumental 
details see the FORS Manual at the ESO web page www.eso.org).
A few additional spectra had already been obtained during the 
commissioning phases of FORS1\&2 in 1998, 1999 and 2000
using the same spectroscopical setup as described above.
All spectra cover a spectral range (in the observer's frame)
from about
3400 \AA \ to about 10000 \AA\ with a spectral scale of 5 \AA /pixel
and a spectral resolution of about 200. 
Although spectra of galaxies as 
faint as I = 26.0 mag have been observed successfully in the FDF, in the
present investigation only galaxies with I\,$ \leq 24.5$\,mag were included.  
Depending on the objects' magnitude and the seeing conditions  
($0\farcs 7$ on average) the integration times ranged between 2 and 12 hours. 
The data reduction (bias subtraction, 
flatfielding, cosmic ray elimination, sky subtraction, 
wavelength calibration, etc.), was performed using standard MIDAS 
routines. A detailed 
description of the FDF spectroscopic observing program 
and the data reduction procedures will be presented by 
Noll et al. (2002).\\
With respect to the photometric redshift catalogue of the FDF (see 
Bender  et al. 2001) our 
spectroscopic sample is complete to about 85\% for our limiting magnitude 
(I\,$ \leq 24.5$\,mag) for photometric redshifts in the range 
$2.2 < z_\mathrm{phot} <  3.5$.
The distribution of the spectroscopic redshifts is in good agreement 
with the photometric redshift 
distribution of the FDF which has peaks at redshifts of around 2.4 and 3.4
(see Fig.~3 in Bender et al. 2001).
Exceptions are the redshifts in the range $1.4 \leq z \leq 2.2$ where 
a lack of 
strong spectroscopic features in our observed wavelength range makes 
a reliable spectroscopic redshift determination 
rather difficult, resulting in an artificial low number
of objects in our spectroscopic sample.

For the present investigation  we selected those FDF galaxies 
showing absorption line spectra with an adequate S/N 
($>10$ per resolution element) for a meaningful quantitative analysis of the \CIV\ resonance doublet.
All these galaxies show typical
starburst characteristics in their spectra such as intense (rest frame) 
UV continua and highly ionized metal absorption lines. 
Three examples of FDF spectra are displayed in Fig.~\ref{fig:exa_spectra}.
The 10 high-redshift galaxies which are dominated by their Ly$_{\alpha}$ 
emission were not included in our study.
This leads to a limitation of the sample to
galaxies with $z < 4$, since most of the few FDF galaxies with 
larger redshifts (up to 5.0) observed so far show strong Ly$\alpha $ 
emission (while pure absorption line spectra dominate at lower redshifts; see 
Noll et al. 2002). 
Also excluded from our study were 4 objects that show absorption lines with
clear emission components, forming P-Cygni profiles. 
(The apparent emission peak redwards from the \CIV\ 
absorption feature in Fig. 1c is not a P Cygni emission component.
The absorption line is unshifted and, as pointed out in the 
caption, the apparent emission component is an artifact).

In order to enlarge our sample somewhat we added
6 additional FORS spectra (matching the criteria listed above) which had been 
observed with the same setup during the FORS commissioning runs
(and are now available from the VLT archive). Four of these additional 
spectra were selected among the gravitationally amplified galaxies 
behind the cluster
1E0657-558 (Mehlert et al. 2001). 
We further included two spectra from the  HDF-S and AXAF Deep 
Field follow-up studies (Cristiani et al. 2000), which met
our criteria.

The comparison sample of IUE low resolution spectra
from the  IUE archive\footnote{\tt http://ines.laeff.esa.es} consists of 
36 local ($z \approx 0$) starburst galaxies investigated by
Heckman et al. (1998)\footnote{
The sample investigated by Heckman et al. (1998) actually contains 45 
galaxies. But 3 of these spectra are not available on the IUE archive,
and 6 were excluded from our study for having either a too low S/N or showing 
strong emission lines or 
defects close to the \SiIV\ and/or \CIV\ lines.}.  All these IUE spectra 
were obtained with the Short Wavelength Prime Camera in the low 
dispersion mode and, therefore, 
cover a similar rest-frame spectral range with a slightly better
spectral resolution
as our FORS spectra of high redshift objects. The IUE spectra were reduced
using the pipeline provided by the archive and smoothed to attain the same 
spectral resolution as the VLT spectra.   

\section{Equivalent widths}
\label{EW}

\begin{figure} 
\includegraphics[width=8.7cm]{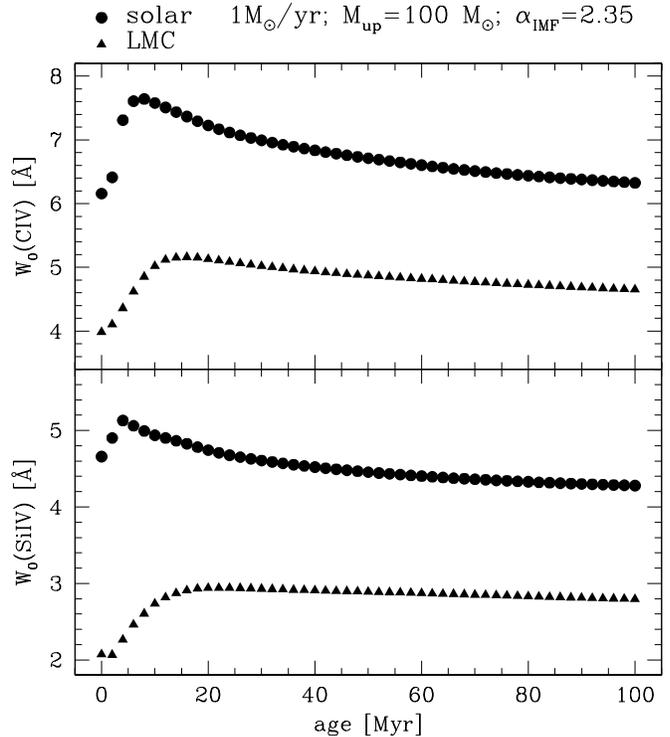} 
\caption{Measured \CIV\ (a) and \SiIV\ (b) equivalent width of the 
synthetic spectra of Leitherer et al. (2001) as a function of the starburst
age.
The model spectra are based on  continuous star formation (1 M$_{\odot}$/yr) 
assuming the model parameter M$_\mathrm{up} = 100$\,M$_{\odot}$ 
and $\alpha_\mathrm{IMF} = 2.35$.
Circles and triangles correspond to solar and LMC metallicity, respectively.
\label{fig:ewmodels} 
} 
\end{figure}

In an earlier investigation, based on
smaller samples of high-redshift galaxies, we noticed an 
apparent anticorrelation between redshift and the strength of 
the \CIV\ 1550 \AA \ doublet (Mehlert et al. 2001, 2002). 
According to Walborn et al. (1995)  
high-excitation lines like \CIV\ and \SiIV\ are produced mostly in stellar 
photospheres and winds and their strengths depend sensitively on the stellar 
metallicity. Although in a few cases a non-negligible ($\leq$\,50\%) 
contamination of
the \CIV\ and \SiIV\ features by interstellar absorption could not be excluded,
Heckman et al. (1998) find for their sample of 45 nearby starburst 
galaxies that the \CIV\ and \SiIV\ absorption
is normally produced by photospheric and stellar wind lines of the
unresolved stars. We found further evidence for a close relation
between the strength of these resonance lines 
and the metallicity of the observed starburst galaxies 
by measuring the equivalent widths 
of the \CIV\ 1550 \AA \ and \SiIV\ 1398 \AA \ doublets 
in synthetic spectra of starburst galaxies with different 
metallicities taken from Leitherer et al. (2001)  
(see Fig.~\ref{fig:ewmodels}). According to Fig. \ref{fig:ewmodels} 
for ages $\ge 10$\,Myr the measured equivalent widths 
depend strongly on the metallicity but are almost independent 
of the age of the starburst.  
Therefore, we measured for those galaxies with 
reliable spectroscopic redshifts $z>1.35$ (i.e. galaxies where 
the \CIV\ doublet was redshifted into our observed spectral range)
the rest-frame equivalent widths $W_{0}$ of this feature, which is defined by
\begin{equation} 
W_0 = \int_{\lambda1}^{\lambda2}\left( 1-\frac{S(\lambda)}{C(\lambda)} \right)d\lambda . 
\end{equation} 
Here we used following approximation:
\begin{eqnarray}
W_0 = \Delta \lambda - \frac{1}{C(\lambda_0)} \int_{\lambda1}^{\lambda2} S(\lambda) d\lambda \,\, \mathrm{with} \nonumber \\
\lambda_1 = \lambda_{0} - \frac{\Delta\lambda}{2};\, \lambda_2 = 
\lambda_{0} + \frac{\Delta\lambda}{2}\, . \nonumber \\
\end{eqnarray} 
The central wavelength for \CIV\ is $\lambda_0$~=~1549.5~\AA\ and  
for the width of the line window we chose $\Delta\lambda = 30$\,\AA.
The continuum flux at the central rest frame wavelength $\lambda_0$ 
was approximated by the mean flux within 
two well defined continuum windows, one on each side of the line window. 
Each of these continuum windows has 
a width of 75~\AA\ and is separated from the line window by 5~\AA .
Since the correct continuum determination is critical for the resulting $W_0$, 
its level was checked interactively. In particular it turned out that the 
influence of the B-Band, that lies in the right continuum window of the 
highest-z objects is negligible with respect to the resulting $W_0$. 

For objects with $z>1.7$, where in addition the \SiIV\ doublet 
($\lambda_0 = 1398.3$~\AA ) became 
visible, we also measured $W_{0}$(\SiIV ) with the same bandwidth and 
continuum window
definitions as for \CIV . 
The same measurements were also carried out in the IUE spectra of the 
comparison sample of $z \approx 0$ starburst galaxies.
Statistical mean errors of the individual $W_{0}$ 
measurements were calculated from the S/N of the individual spectra
and the errors of the continuum fits ($\leq 10$\,\%). Equivalent width
measurements in the spectra of faint objects can be affected severely
by errors in the sky background subtraction. Therefore, during the reduction of
our FORS spectra a particular effort was made to keep these errors low. 
Various tests showed that for all FORS spectra used in this study the
errors in determining the continuum level remained below 
5\% (see Noll et al. 2002). Hence, systematic errors in the measured 
equivalent widths due to an incorrect sky subtraction are 
well below the statistical errors in most cases.

Another source of systematic errors in our $W_{0}$ measurements
is our low spectral resolution, which in most cases did not allow
us to resolve the line profiles. However, since equivalent widths
measurements
of strong isolated lines are in principle independent of the spectral
resolution and since all our conclusions are based on differences
between measurements carried out with the same procedure  
in spectra of the same resolution, these systematic errors are
expected to cancel out and therefore are not expected to affect the
results of this paper significantly. 
On the other hand, since the \CIV\ doublet is not a truly isolated
feature, our numerical results for the equivalent widths 
should not be directly compared to results obtained from 
spectra with a different spectral resolution.    
 
\begin{figure} 
\includegraphics[width=8.7cm]{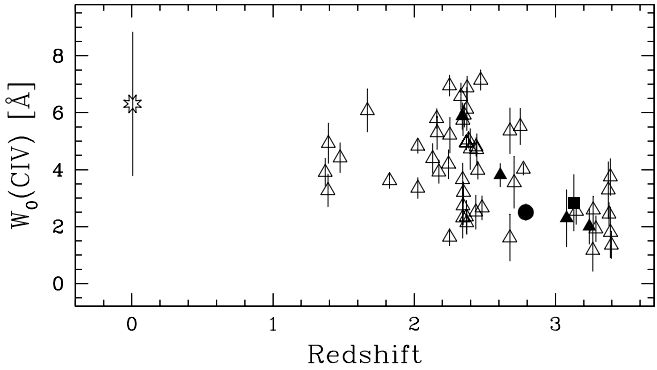} 
\caption{Measured \CIV~$\lambda$ 1550 rest-frame equivalent widths
as a function of redshift. The various symbols 
have the following meaning: Open star at $z\approx 0$: Average and 
$1\sigma$\,rms
scatter for the 36 (Heckman et al. 1998) local starburst galaxies.
Open triangles: FDF galaxies. Filled triangles: 
Galaxies in the field of the cluster 1E0657-558 (Mehlert et al. 2001). 
Filled circle and
square: Galaxies in the HDF-S and the AXAF Deep Field, respectively 
(Cristiani et al. 2000). 
\label{fig:civz} 
} 
\end{figure}

The results of our $W_{0}(\CIV)$ measurements 
are listed in Table~\ref{tab:ew} and plotted
in Fig.~\ref{fig:civz}. In order to avoid a crowding 
of data points at $z \approx 0$, we plotted 
for the local (IUE) starburst galaxies only the average value and indicate
the $1\sigma$ scatter of the individual values by a bar. 
For the high-redshift galaxies the individual
data points and their mean errors are given.   
\begin{table*}
\caption{Measured \CIV\ and \SiIV\ rest-frame equivalent widths for the 
57 high-z galaxies. The absolute B-magnitudes derived and discussed in Sect.~\ref{luminosity} are 
also listed.  
}
\label{tab:ew}
\vspace{0.5cm} 
\centerline{
\begin{tabular}{|l|c|c|c|c|c|c|c|}
\hline
  No.  &  $z$ &  I &   $W_{0}(\CIV )$ & d$W_{0}(\CIV )$  &   $W_{0}(\SiIV )$ & d$W_{0}(\SiIV )$  & M$_\mathrm{B}$\\
 & & $[$mag$]$ & $[$\AA$]$ & $[$\AA$]$ & $[$\AA$]$ & $[$\AA$]$ & $[$mag$]$ \\
\hline
FDF-1208   &    2.18  &     23.68   &    3.92   &   0.41  &     3.31   &   0.44  &  -22.45  \\    
FDF-1331   &     3.39  &     23.89   &    1.80   &   0.90  &     --    &     --  &  -23.30  \\    
FDF-1555   &    3.26  &     23.88   &    1.16   &   0.74  &     1.29   &   0.69  &  -22.36  \\    
FDF-1578   &    2.71  &     24.25   &    3.545   &   0.90  &    2.97   &   0.81  &  -21.98  \\    
FDF-1691   &    2.34  &     23.89   &    3.65   &   0.57  &     4.30   &  0.60   &  -21.53  \\    
FDF-1709   &    1.67  &     24.33   &     6.07   &   0.75  &     --    &     --  &  -20.46  \\    
FDF-1744   &     2.37  &      24.10   &    2.34   &   0.58  &   4.27   &   0.56  &  -22.21  \\    
FDF-1922   &    1.83  &     23.36   &     3.62   &    0.28  &   0.95   &   0.38  &  -21.64  \\    
FDF-2033   &    2.75  &     24.08   &    5.51   &   0.65  &     4.05   &   0.52  &  -21.62  \\    
FDF-2274   &    2.25  &     23.34   &    1.62   &   0.31  &     2.66   &   0.33  &  -21.66  \\    
FDF-2418   &     2.33  &     23.16   &    6.57   &   0.47  &    6.36   &   0.46  &  -23.19  \\    
FDF-2495   &     2.45  &     23.31   &    3.98   &   0.32  &    3.24   &   0.32  &  -22.32  \\    
FDF-2636   &    2.25  &     23.43   &    5.21   &   0.63  &     6.41   &   0.70  &  -22.77  \\    
FDF-3005   &     2.25  &     23.51   &    6.95   &   0.38  &    8.18   &   0.43  &  -22.63  \\    
FDF-3163   &    2.44  &     23.35   &    4.80   &   0.33  &     5.45   &   0.33  &  -22.97  \\    
FDF-3173   &    3.27  &     23.91   &    2.59   &   0.48  &     4.24   &   0.44  &  -22.51  \\    
FDF-3300   &    2.37  &     23.91   &    2.14   &   0.42  &     2.16   &   0.41  &  -21.79  \\    
FDF-3374   &    2.38  &     23.34   &    5.05   &   0.30  &     5.21   &   0.27  &  -22.65  \\    
FDF-3810   &    2.37  &     22.67   &    4.95   &   0.25  &     5.59   &   0.26  &  -23.18  \\    
FDF-3874   &    2.48  &      23.30   &    2.66   &   0.43  &    3.83   &   0.43  &  -23.15  \\    
FDF-3875   &    2.24  &     24.53   &    4.19   &   0.51  &     3.38   &   0.52  &  -20.73  \\    
FDF-3958   &    2.13  &     23.87   &    4.40   &   0.53  &     1.43   &   0.57  &  -20.98  \\    
FDF-3999   &    3.39  &     24.00   &    3.74   &   0.65  &     7.89   &   0.56  &  -22.70  \\    
FDF-4049   &    1.48  &     23.00   &    4.42   &   0.53  &       --   &     --  &  -21.76  \\    
FDF-4795   &    2.16  &     23.31   &    5.80   &   0.36  &     5.93   &   0.39  &  -22.35  \\    
FDF-4871   &    2.47  &     23.39   &    7.14   &    0.35  &    6.05   &   0.34  &  -22.62  \\    
FDF-4996   &    2.03  &     23.25   &    3.35   &   0.37  &     1.19   &   0.44  &  -21.77  \\    
FDF-5058   &    2.03  &     23.34   &    4.82   &   0.25  &     3.40   &   0.27  &  -21.57  \\    
FDF-5072   &    1.39  &     22.45   &    3.27   &   0.57  &       --   &     --  &  -21.88  \\    
FDF-5135   &    2.34  &     23.62   &    2.31   &   0.71  &     1.99   &   0.77  &  -22.73  \\    
FDF-5152   &    1.37  &     22.65   &     3.91   &   0.50  &      --   &     --  &  -21.55  \\    
FDF-5165   &    2.35  &     23.26   &     5.73   &   0.55  &    6.02   &   0.53  &  -23.02  \\    
FDF-5190   &    2.35  &     24.39   &     2.73   &   0.68  &    2.65   &   0.64  &  -22.74  \\    
FDF-5215   &    3.15  &     22.98   &    2.54   &   0.47  &     2.45   &   0.40  &  -23.18  \\    
FDF-5227   &    2.40  &     23.85   &    4.73   &    0.72  &    2.02   &   0.79  &  -21.79  \\    
FDF-5504   &    3.38  &     23.63   &    3.29   &   1.00  &     7.19   &   0.73  &  -23.65  \\    
FDF-5550   &     3.38  &     23.12   &    2.44   &   0.41  &    5.23   &   0.31  &  -23.23  \\    
FDF-5903   &    2.77  &     22.33   &    4.02   &   0.21  &     4.32   &   0.16  &  -23.23  \\    
FDF-6024   &    2.37  &     22.00   &    4.93   &   0.20  &     5.71   &   0.19  &  -23.34  \\    
FDF-6063   &    3.40  &     22.56   &    1.36   &   0.49  &     4.24   &   0.42  &  -23.35  \\    
FDF-6069   &    2.68  &     24.22   &    5.36   &   0.81  &     3.45   &   0.70  &  -21.74  \\    
FDF-6287   &    2.68  &     24.11   &    1.61   &   0.82  &       --   &    --   &  -22.04  \\    
FDF-6372   &    2.35  &     23.38   &    3.20   &   0.37  &     3.43   &   0.35  &  -22.18  \\    
FDF-6407   &    2.16  &     23.59   &    5.29   &    0.59  &    4.26   &   0.72  &  -22.31  \\    
FDF-6864   &    1.39  &     23.41   &    4.93   &   0.71  &       --   &    --   &  -20.87  \\    
FDF-6934   &    2.44  &      22.90   &    4.71   &   0.56  &    3.01   &   0.60  &  -23.05  \\    
FDF-6947   &    2.36  &     23.83   &    5.91   &   0.42  &     4.90   &   0.45  &  -21.94  \\    
FDF-7029   &    2.37  &     23.63   &    6.88   &   0.42  &     5.82   &   0.42  &  -23.07  \\    
FDF-7307   &    2.44  &     24.07   &    2.51   &   0.60  &     3.05   &   0.62  &  -21.20  \\    
FDF-7342   &    2.37  &      23.80   &    6.13   &   0.76  &    4.58   &   0.75  &  -22.06  \\    
FDF-7539   &    3.29  &     23.51   &    1.92   &   0.46  &     4.22   &   0.41  &  -22.41  \\    
\hline
ES0657-A    &  2.34     & 24.50   &        5.88    &       0.48     &      3.98    &    0.45 & -- \\
ES0657-C    &  3.08     &  24.89  &        2.30    &        1.00     &      7.07    &    0.62 & -- \\
ES0657-J    &  2.61     & 22.98   &        3.81    &       0.42     &     2.41    &    0.26  & -- \\
ES0657-Core &  3.24   &  24.31  &         2.01    &       0.63     &     3.45   &   0.53 & -- \\
HDFS-047    &  2.79     & -- &     2.50    &       0.30     &      4.40    &       0.17	& -- \\
AXAF-028    &  3.13     & -- &        2.84    &        1.00     &      3.45    &       0.97  & -- \\
\hline
\end{tabular}
}
\end{table*}
\begin{figure} 
\includegraphics[width=8.7cm]{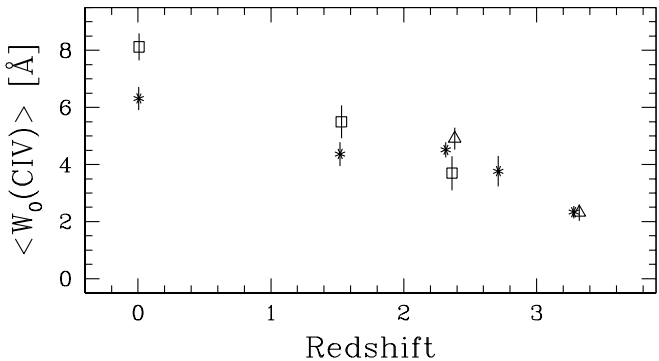} 
\caption{Averages of the measured \CIV~$\lambda$ 1550 rest frame 
equivalent widths within selected redshift bins (see Table~\ref{tab:mciv}) 
as a function of redshift for all galaxies shown in 
Fig.~\ref{fig:civz} (asterisks). 
The bars indicate the mean errors of the averages. 
Open triangles show all FDF galaxies brighter than 
M$_\mathrm{B} = -22.28\,$mag, open squares all FDF galaxies with 
$-21.52\,\mathrm{mag} \leq \mathrm{M}_\mathrm{B} \leq -20.38\,\mathrm{mag}$
(see Sect.~\ref{luminosity} for discussion).
\label{fig:mcivz_fors} 
} 
\end{figure} 
\begin{table}
\caption{Averages and mean errors of the measured \CIV~$\lambda$ 1550 rest 
frame 
equivalent widths within selected redshift bins for all galaxies listed 
in Table~\ref{tab:ew}. N is the number of objects within each bin.  
}
\label{tab:mciv}
\vspace{0.5cm} 
\centerline{%
\begin{tabular}{|l|c|c|c|r|}
\hline
       $z$ interval &  $<z>$  &  $<\CIV >$ &  m.e.(\CIV ) &  N \\  
 & & $[$\AA$]$ & $[$\AA$]$ & \\	       				     
\hline	       				     
0.00 -- 0.02  &   0.01   &   6.31    &    0.41   &    36  \\
1.00 -- 1.99   &   1.52   &   4.37    &    0.41   &     6  \\
2.00 -- 2.49   &   2.32   &   4.51    &   0.27   &    32  \\
2.50 -- 2.99   &   2.71   &   3.77    &   0.53   &    7  \\
 $\geq 3.00$  &   3.28   &   2.33    &   0.22   &     12  \\
\hline
\end{tabular}
}
\end{table}

As demonstrated by Fig.~\ref{fig:civz} our high-redshift galaxies 
with $z<2.5$ show about the same average \CIV\ equivalent widths and
about the same scatter around the average as the local starburst
galaxies. However, for redshifts larger than about 2.5 the average
\CIV\ equivalent widths and their scatter clearly decrease with $z$
in our sample. As described in Sect.~\ref{Sample}
this decrease is not driven by any selection effect 
since the spectroscopic redshift distribution of the included FDF 
galaxies is in good agreement with the photometric redshift distribution.

Fig.~\ref{fig:civz} obviously
confirms the effect suspected by Steidel et al. (1996a) quantitatively.
In order to estimate (in view of the observed scatter) the statistical 
significance of the effect, we calculated averages and their mean errors  
of the $W_{0}$(\CIV ) values for selected redshift bins. The results are
listed in  Table~\ref{tab:mciv} and plotted in 
Fig.~\ref{fig:mcivz_fors}. The table confirms that there is no 
statistically significant difference between the results for the 
first three bins while the difference between the local sample and 
our starburst galaxies with $z>3.0$ is highly 
significant ($>9 \sigma$). Tests with other bin sizes and binning intervals
showed that the high significance of the result persists for any reasonable  
bin distribution. 

\begin{figure} 
\includegraphics[width=8.7cm]{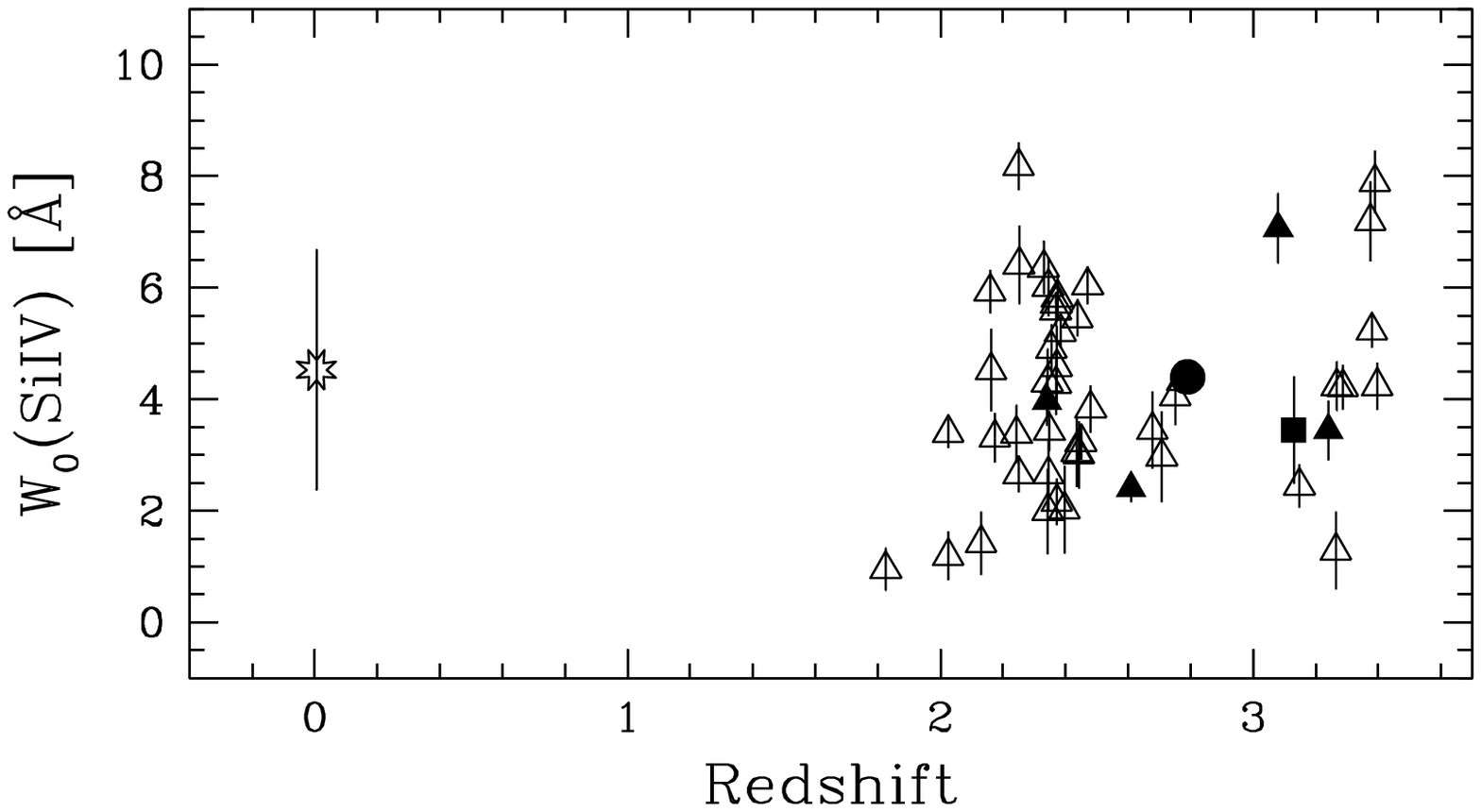} 
\caption{Measured \SiIV~$\lambda$ 1400 rest-frame equivalent width
as a function of redshift. Symbols as is Fig.~\ref{fig:civz}.
\label{fig:siivz} 
} 
\end{figure}

\begin{figure} 
\includegraphics[width=8.7cm]{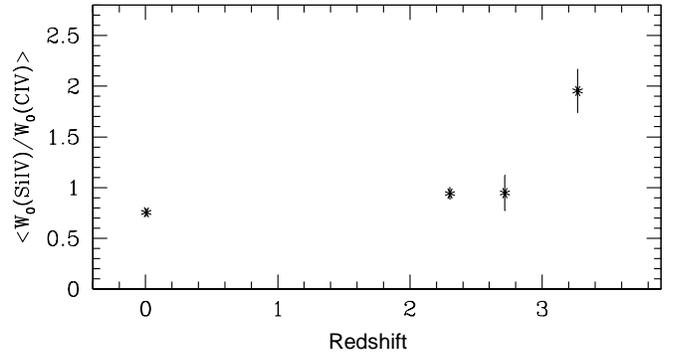} 
\caption{Averages of the ratios between the measured \SiIV\ and \CIV\ rest 
frame equivalent widths as a function of redshift for all galaxies shown in 
Fig.~\ref{fig:siivz}. The mean error of each average, calculated from the 
scatter of the individual values, is indicated by a bar.  
\label{fig:mratioz} 
} 
\end{figure}

In Fig.~\ref{fig:siivz} we present analogously to Fig.~\ref{fig:civz} 
the observed \SiIV\ equivalent width values as a function of redshift.
Similarly as in the case of the \CIV\ doublet the average \SiIV\ 
strength does not 
change with redshift for $z<2.5$. However, unlike the average \CIV\
strength the average \SiIV\ 
$W_{0}$ values remain at the local value even beyond $z>2.5$. As a result,
the ratio between the \SiIV\ and \CIV\ resonance doublets, which is
practically constant for low $z$, varies for high redshifts in our sample.
This is demonstrated quantitatively by Table~\ref{tab:mratio} and Fig.~\ref{fig:mratioz}.    

As noted above, for local starburst galaxies the \CIV\ doublet as well as
the \SiIV\ doublet are good metallicity indicators. Since both elements are
produced during the evolution and explosion of massive stars, a greatly
different relative chemical abundance of C and Si in the high-$z$
starburst galaxies appears very unlikely. In spite of the
unsatisfactory state of present SN II models and the remaining large
uncertainties concerning the intermediate mass element yields for
different initial stellar masses, it seems very difficult to enhance
Si production relative to C. On the other hand, it is well known that
individual hot stars of the same metallicity show a wide range of  
\CIV\ to \SiIV\ line ratios.  
Strong \CIV\ absorption is well known to be 
universally present in O~stars of all luminosity classes, 
while $W_{0}(\SiIV)$ is luminosity dependent and decreases 
rapidly from supergiant stars to dwarfs (Walborn \& Panek~1984, 
Pauldrach et al.~1990, Leitherer et al. 1995).
This line is, therefore, used as a luminosity indicator in UV stellar
classification schemes. Moreover, \SiIV\ has a pronounced maximum 
in early B stars
while \CIV\ changes monotonically with temperature.  
Finally the \SiIV\ strength is more strongly affected by population 
differences (i.e. stellar age differences) than the \CIV\ doublet.
A scatter in these population differences can easily mask any 
metallicity dependence in the \SiIV\ line strength. 

Hence, assuming that the observed absorption lines are dominated by 
contributions of the stellar photospheres and winds, 
the $W_{0}(\SiIV)/W_{0}(\CIV)$ ratio can, in 
principle, be used to derive informations on the star formation history 
(instantaneous or continuous) and/or the stellar mass distribution
(see also Mas-Hesse \& Kunth 1991).
Therefore, Fig.~\ref{fig:mratioz} can possibly be  
understood by assuming that at the epochs corresponding to $z>3$ (i.e.
during the first two Gyrs of the universe) instantaneous star
bursts played an important role while ``continuous star formation''
is the normal mode for local and lower redshift starburst galaxies.
However, since the different parameters which determine the 
value of $W_{0}(\SiIV)/W_{0}(\CIV)$ cannot be disentangled 
reliably without including additional lines in the analysis, 
spectra of higher resolution and higher S/N are required to finally settle 
this issue.

The investigation of the purely interstellar lines of lower ionisation
like e.g. \SiII~$\lambda$ 1260, \OI /\SiII~$\lambda$ 1303 and 
\CII~$\lambda$ 1335 is in progress and the results will be published in a 
separate paper. 

\begin{table}
\caption{Average and mean error of the ratios between the measured 
\SiIV\ and \CIV\ rest 
frame equivalent widths 
within selected redshift bins for all high-z galaxies for which 
$W_0(\SiIV )$ has been measured. N is the number of galaxies within each bin.   
}
\label{tab:mratio}
\vspace{0.5cm} 
\centerline{%
\begin{tabular}{|l|c|c|r|}
\hline
         $<z>$  &  $<\SiIV /\CIV >$ &  m.e.(\SiIV /\CIV ) &  N \\  
\hline	       				     
       0.00   & 0.76   & 0.04   &    36  \\
       2.30   & 0.94   & 0.06   &    33  \\
       2.72   & 0.95   & 0.18   &    6  \\
       3.27   & 1.95   & 0.22   &    11  \\
\hline
\end{tabular}
}
\end{table}

\section{Metallicities}
\label{metallicities}

Since, in contrast to the variations of the \SiIV\  line, differences 
of the \CIV\ line strength cannot be easily explained by population
differences in the starburst galaxies, the observed decrease of the \CIV\
equivalent width values for $z>2.5$ in our sample can at present only
be interpreted as a metallicity effect. Hence, the decrease of
$W_{0}$(\CIV) with $z$ is expected to contain information on the 
evolution of the metal content of starburst galaxies with cosmic age.
In order to derive a more quantitative measure of the metallicity evolution 
apparently observed in Fig.~\ref{fig:civz}, we made an attempt to calibrate 
the observed $W_{0}(\CIV)$ values in terms of the O/H ratios. For this
purpose we used the oxygen abundances listed in  Heckman et al. (1998)
for all the local starburst galaxies of this sample and derived metallicities
using the relation $\log Z = 12 + \log (O/H)$.

\begin{figure}
\includegraphics[width=8.7cm]{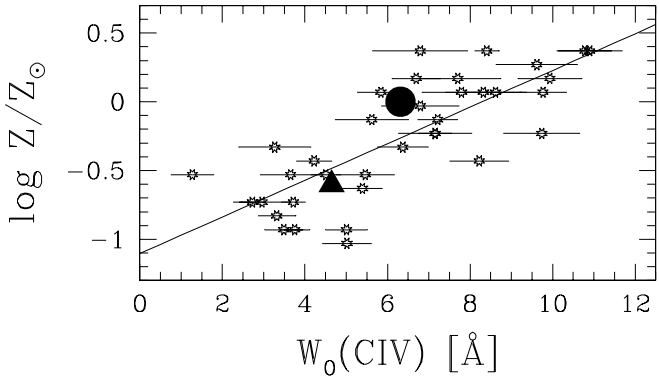} 
\caption{Metallicity $\log Z/Z_{\odot}$ in terms of oxygen abundance 
of 36 local starburst galaxies (from Heckman et al. 1998) as a function
of the measured \CIV~$\lambda$ 1550 equivalent width (open stars). 
The typical error for $\log Z/Z_{\odot}$ is $\leq$ 0.1 dex.
The solid line
gives the best linear fit to the data.
For comparison the $W_{0}(\CIV )$ of synthetic starburst spectra (from 
Leitherer et al. 2001; see Fig.~\ref{fig:ewmodels}) 
at an age of 100 Myr for solar (filled circle)  
and LMC (0.25 solar, filled triangle) metallicity 
have also been included in the figure.   
\label{fig:metalcalib} 
} 
\end{figure} 
In Fig.~\ref{fig:metalcalib} we plotted for the local starburst galaxies
the metallicity relative 
to the solar value ($\log Z_{\odot} = 8.93$)
as a function of our measured $W_{0}(\CIV )$ values.
We also included the theoretical values determined for
synthetic starburst galaxies with solar and LMC metallicities
taken from Leitherer et al. (2001). Although the scatter is 
rather large (RMS = 0.27), the plot indicates a dependence of log Z on 
$W_{0}(\CIV )$ which can be approximated by a linear relation.
The best linear least square fit to these data gives
\begin{equation}
\log Z/Z_{\odot} = 0.13 (\pm 0.02) \times W_{0}(\CIV ) -1.10 (\pm 0.12).
\label{eq:metalcalib} 
\end{equation}
This calibration of the \CIV\ strength in terms of metallicity
should be a reasonable approximation for statistical applications at least
for local starburst galaxies and the objects with $z<2.5$.
However, in view of the population differences evident from the
different \SiIV\ to \CIV\ ratio, its applicability to the $z>2.5$
objects is less clear. 
Nevertheless, because of the absence of
other more reliable calibration procedures, 
and since (in view of the physics of hot stars)  the relation between the
\CIV\ strength and the metallicity should not be much affected by 
population details of starbursts
we will assume for the following 
that the correlation between the oxygen abundances and  
$W(\CIV )$ observed for low redshift starburst galaxies 
is also valid at high redshifts. With this assumption we 
convert our observed \CIV\ equivalent width values 
to metallicities using Eq.~\ref{eq:metalcalib}. 
In this way we obtain for our starburst galaxies with $z>3$ 
($<z> = 3.24$) an average metallicity of  
about $0.16\,Z_{\odot}$ and for $<z> = 2.34$ a value of 
$0.42\,Z_{\odot }$. The corresponding 
local ($z = 0$) value would be 0.56 $Z_{\odot }$. 
In terms of cosmic time scales (for a universe with $\Omega _{\Lambda }
= 0.7, \Omega _{M} =0.3, H_{0}$ = 67 km sec$^{-1}$ Mpc$^{-1}$ used throughout the paper) this would 
correspond to an increase of
the mean metallicity in starburst galaxies
by a factor of 2.5 within $\approx 1$\,Gyr between cosmic ages of about
1.9 Gyrs and 2.9 Gyrs.
For later epochs the data suggest only little further
enrichment. Because of the approximative nature of Eq.~\ref{eq:metalcalib} 
these numbers are rough estimates only.
Still, they agree surprisingly well 
with earlier theoretical predictions of the cosmic
chemical enrichment history of the universe
by e.g. Fritze-von Alvensleben (1998) and 
by Renzini (1998, 2000), who predicts 
that the metallicity had been $\approx 0.1\,Z_{\odot}$ at $z = 3$ and has 
increased to a value of $\approx 1/3\,Z_{\odot}$ in the local universe. 

Two effects may affect the validity of Eq.~\ref{eq:metalcalib} at high 
redshifts: First, in the
IUE spectra of the local starburst galaxies contributions of
the Milky Way halo components are present, while they are absent in the 
high-z spectra. Savage \& Massa (1987) showed that the \CIV\
equivalent width of distant halo
stars are normally $< 0.5$~\AA . This is one order of magnitude 
lower than the values measured in the local starburst galaxies.
Hence its effect on Equ. 3 should be on a 10\,\% level, at most.
Secondly, the spectrum of the magnified high-$z$ object MS1512-cb58
($z$ = 2.727) published by Pettini et al. (2000)
suggests that the contribution of the interstellar
line to the \CIV\ absorption feature increases with redshift. 
If this holds for all galaxies at similar redshifts,
this could result in a general difference of the mean \CIV\ equivalent
width between local starburst galaxies and high-redshift objects,
but it could not explain the observed evolution of the \CIV\
equivalent width between $z$ = 2.5 and $z$ = 3.5.
Furthermore, by applying the local relation to high redshifts we 
would overestimate the mean metallicity at young epochs.

\section{Luminosity effects}
\label{luminosity}
\begin{figure} 
\includegraphics[width=8.7cm]{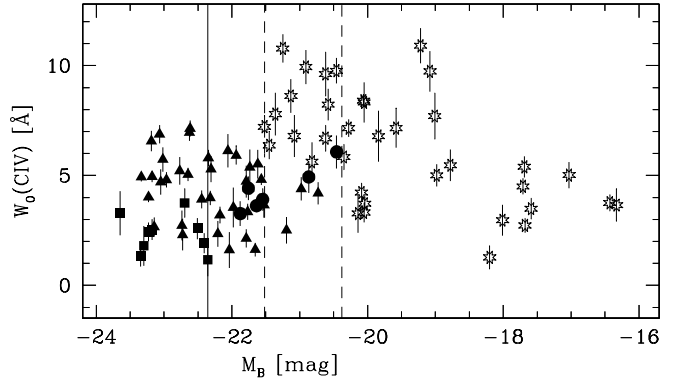} 
\caption{ 
Measured \CIV~$\lambda$ 1550 rest-frame equivalent widths of the 
local starburst galaxies (open stars) and the high-$z$ FDF (filled symbols) 
versus the absolute B-magnitude. For the latter ones we indicated objects
within the different redshift bins
$ 1 \leq z < 2$, $ 2 \leq z < 3$ and $ z \geq 3$ by filled circles, triangles and squares, respectively.
The solid vertical line indicates M$_\mathrm{B} = -22.36\,$mag , while the dashed vertical lines indicate 
M$_\mathrm{B} = -21.52\,$mag and M$_\mathrm{B} = -20.38\,$mag
\label{fig:civmag} 
} 
\end{figure} 
When interpreting the data described above one has to keep in mind
that for local ($z \approx 0$) galaxies the metallicities are known
to depend on the galaxies' blue and infrared luminosities, with luminous galaxies
tending to have higher metallicities (see e.g. Kobulnicky \& 
Zaritsky 1999 and Heckman et al. 1998). Since at high redshifts we observe only very luminous galaxies. Therefore, if 
a metallicity-luminosity
correlation exists these bright objects should be metal rich and we
should find an opposite correlation between metallicity and redshift
than the one detected in this work. 
To test whether at high redshifts a metallicity-luminosity
correlation does exist and may affect our detected metallicity evolution
we plotted in Fig.~\ref{fig:civmag} 
the absolute B-magnitudes M$_\mathrm{B}$ of 
all high-redshift FDF galaxies as well as M$_\mathrm{B}$ for the local starburst galaxies. 
For the local objects the M$_\mathrm{B}$ was taken from Heckman et al. 1998 
(transformed to the cosmology used in this paper), 
for the FDF galaxies we
computed M$_\mathrm{B}$ as follows: We derived the best fitting SED, scaled to the total I flux
derived by SExtractor (FLUX\_AUTO) as determined by our
photometric redshift code (see Bender et al. 2001). 
Then this SED was transformed to $z=0$ (using the observed spectroscopic
redshift) to derive the rest-frame B-magnitude of the galaxies. Since
for the redshift range in question the measured J and K bands
bracket the rest-frame B, this procedure is nearly equivalent to an
interpolation, minimizing the uncertainties in the K corrections.  A
detailed description of the method can be found in Gabasch et
al. (2002). Using the photometric instead of the spectroscopic redshifts
would produce a typical
variations of $\pm 0.2$ mag. Absolute magnitudes were derived assuming the
cosmology parameter $H_0=67$, $\Omega_m=0.3$, $\Omega_{\Lambda}=0.7$. 
Our M$_\mathrm{B}$ have been corrected for foreground Galactic extinction but not for any internal extinction in the starburst galaxies.
From Fig.\ref{fig:civmag} we see that the local starburst galaxies indeed show the expected 
correlation between $W_{0}(\CIV )$ and the luminosity. On the other hand, for 
the high-redshift galaxies we cannot determine whether a metallicity-luminosity relation 
does exist or not, since we do not have any faint objects in our high-$z$ 
sample. But it is evident that the high-redshift galaxies
are on average overluminous for their 
metallicities compared with local starburst galaxies.
This agrees well with earlier results from
Pettini et al. (2001) and Kobulnicky \& Koo (2000) who find this trend for Lyman-break galaxies.
Hence, if a metallicity-luminosity relation does exist at high redshifts,
our data suggest that it has a clear offset to the local correlation, 
which seems to evolve with redshift.
Moreover, from Fig.\ref{fig:civmag}
it is obvious that for the high-redshift galaxies there is no correlation between the 
measured $W_{0}(\CIV )$ and the luminosity that could cause
the correlation with $z$ found in this paper.

Since at high redshifts we do not have any faint objects in our sample, while 
in the local universe we do not find bright starburst galaxies,
we have to make sure that our detected metallicity evolution with redshift
is not produced by comparing different objects at different redshifts. 
For that reason we separately investigated all galaxies, which are brighter than the faintest one at 
$z \geq 3$ (which is M$_\mathrm{B} = -22.36\,$mag; solid line in Fig.~\ref{fig:civmag}). 
In our sample we
only find galaxies brighter than this limit for
$z \geq 2$. Their
average values of the measured $W_{0}(\CIV )$
and the mean error at redshift 2.4 and 3.3 as well as the
single $W_{0}(\CIV )$ measurement for this brightest local galaxy
are additionally 
indicated by open triangles
in Fig.\ref{fig:mcivz_fors}. Obviously these subsample
show the same trend with decreasing redshift as 
the total galaxy sample at $z\geq 2$.
Furthermore we investigated all galaxies fainter than M$_\mathrm{B} = -21.52\,$mag 
(brightest local galaxy) and brighter than M$_\mathrm{B} = -20.38\,$mag 
(faintest galaxy with $z\geq 1$). The average values of the measured $W_{0}(\CIV )$
and the mean error at redshift 0, 1.5 and 2.4  are also indicated in 
Fig.\ref{fig:mcivz_fors} by open squares and again show the same 
trend with decreasing 
redshift. From these test we
conclude that the observed dependence of $W_{0}(\CIV )$ on redshift 
is not caused by a 
luminosity effect.
Moreover the two open symbols in Fig.~\ref{fig:mcivz_fors} for 
$z \approx 2.4$  
indicate that a metallicity-luminosity correlation also exists at this 
redshift.

\begin{figure} 
\includegraphics[width=8.7cm]{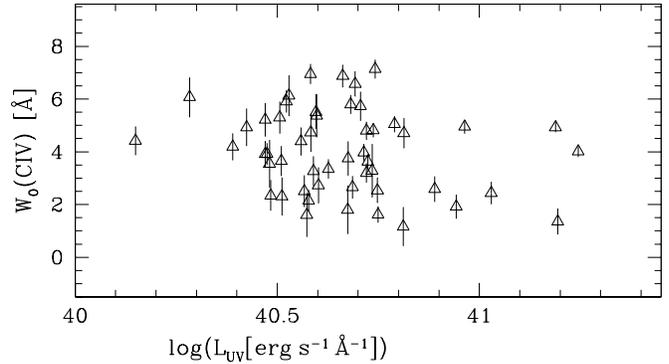} 
\caption{ 
Measured \CIV~$\lambda$ 1550 rest-frame equivalent widths of the 
high-$z$ FDF galaxies 
as a function of the UV luminosity L$_\mathrm{UV}$, derived from the
flux between 1432\,\AA\ and 1532\,\AA\ (as defined by 
Kinney et al. 1993)
\label{fig:civlum} 
} 
\end{figure}

The following additional selection effects could be present 
(and possibly weaken) the observed 
correlation between metallicity and redshift at high-$z$: 
It could, in principle, be possible that at high-$z$ we preferentially see 
objects with low internal extinction, which have low dust content and hence low
metallicity. In this case we would expect to find a negative correlation 
between the UV luminosity of our galaxies and their metallicity.
To test whether this correlation is present in our high-$z$ galaxies we 
calculated the UV luminosity as follows:
\begin{eqnarray} 
\log (\mathrm{L}_\mathrm{UV}) = \log <F>_{1432}^{1532} + \frac{(\mathrm{m-M})}{2.5} 
\nonumber \\
+ \log (4\,\pi (10pc)^2)  \nonumber \\ 
\end{eqnarray} 
where $<$\,$F$\,$>_{1432}^{1532}$ is the mean flux between 1432\,\AA\ and 
1532\,\AA\ (as defined by Kinney et al. 1993), (m-M) is the distance modulus 
calculated using the cosmological parameters mentioned above. 
We did not correct for intrinsic reddening since such a correction 
would have involved considerable uncertainties and is not needed for the
the present test.
From Fig.~\ref{fig:civlum}, where we plotted  
the \CIV\ equivalent widths for the 
high-$z$ FDF galaxies as a function of L$_\mathrm{UV}$, we see that no such 
correlation is evident for our high-$z$ objects. Hence our galaxy sample seems not to be affected by this
selection effect.

\section{Comparison with literature data}
\label{litcomparison}

\begin{figure} 
\includegraphics[width=8.7cm]{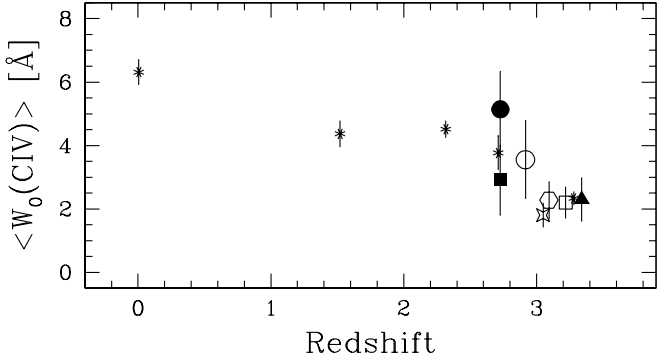} 
\caption{Comparison of the observed \CIV~$\lambda$ 1550 rest frame 
equivalent widths as a function of redshift for our
galaxy sample (asterisks) with the
mean values of $W_0(\CIV )$ derived from high-z
galaxy spectra from Pettini et al. 1998 (P98; open circle), Steidel et al. 
1996a (S96; open square), Steidel et al. 98 (S98; open hexagons) and 
Lowenthal et al. (L98; open diamond). The bars denote mean errors.
Single measurements from
galaxies observed by Pettini et al. 2000 (P00; filled circle), 
Yee et al. 1996 (Y96; filled square) and Trager et al. 1997 
(T97; filled triangle) are also shown. 
Note that the Y96 and P00 observed the same object namely the well known 
lensed galaxy MS1512-cb58.
\label{fig:mcivz_fors_lit} 
} 
\end{figure} 
\begin{table}
\caption{Line 1 -- 4: Average of the measured \CIV~$\lambda$ 1550 rest-frame 
equivalent widths for high-z galaxies observed by Pettini et al. (1998; P98),
Lowenthal et al. (1997; L97) and Steidel et al. (1996a \& 1998; S96 \& S98). 
Line 5 -- 7: Measured \CIV~$\lambda$ 1550 rest frame equivalent width for MS1512-cb58
observed by Yee et al. (1996; Y96) and Pettini at al. (2000; P00) and for
object DG-433 observed by Trager et al. (1997; T97).
A * indicates that the equivalent widths were measured using 
the same method applied to the the high-z galaxies presented in this work.
A \dag\ indicates that the equivalent widths were measured from enlarged 
tracings.
}
\label{tab:mciv_lit}
\vspace{0.5cm} 
\centerline{%
\begin{tabular}{|l|c|c|c|c|}
\hline
 $<z>$  &  $<\CIV >$ &  m.e.(\CIV ) &  N & Reference\\  
 & $[$\AA$]$ & $[$\AA$]$ & & \\	       				     
\hline	       				     
 2.92   &  3.55   &    1.23   &   5  & P98*\\
 3.05   &  1.80   &    0.39   &   5  & L97$^{\dag}$ \\
 3.09   &  2.28   &    0.59   &   4  & S98$^{\dag}$ \\
 3.22   &  2.20   &    0.50   &   2  & S96$^{\dag}$ \\
\hline
 $<z>$  &  $W_{0}(\CIV )$  &  d$W_{0}(\CIV )$ ) &  N & Reference\\  
 & $[$\AA$]$ & $[$\AA$]$ & & \\	       				     
\hline	       				     
 2.73   &  5.14   &    1.20   &   1  & P00*\\
 2.73   &  2.91   &    1.11   &   1  & Y96*\\
 3.34   &  2.30   &    0.70   &   1  & T97$^{\dag}$ \\
\hline
\end{tabular}
}
\end{table}
\begin{figure} 
\includegraphics[width=8.7cm]{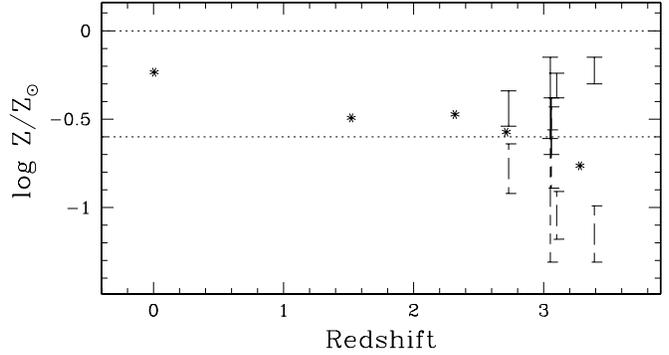} 
\caption{Mean of the estimated metallicities
 as a function of redshift for all galaxies shown in 
Fig.~\ref{fig:civz}. 
The oxygen abundances and the statistical uncertainties
derived from $R_{23}$ ratio for 5 Lyman-break galaxies 
investigated by Pettini et al. (2001) and Teplitz et al. (2000) are 
indicated by the vertical lines: solid line: upper branch results; 
dashed line: lower branch result. 
Solar and LMC metallicities are indicated by the 
horizontal dotted lines. 
\label{fig:mmet_z_lit} 
} 
\end{figure} 
Although most published spectra of high redshift galaxies lack the S/N
required to carry out a study of the type presented here we tried to 
compare 
our results with the limited information available on this subject in 
the literature.   
For this purpose we measured the \CIV\ equivalent width for all 
high-z galaxy spectra published in the papers listed in the caption of 
Table~\ref{tab:mciv_lit}. As in our samples, spectra with strong Ly$\alpha $
emission were disregarded, and only equivalent width values with estimated
(rest frame) mean errors $\leq 1.0$\,\AA \ were used for the comparison.
The number of the (for our purpose) usable spectra of each 
publication is listed in column 4 of Table~\ref{tab:mciv_lit}. 
(Tests showed that including less accurate data lead to similar results,
but with much larger statistical errors).   
For the objects observed by Yee et al. (1996) and by Pettini 
et al. (1998, 2000) the author kindly made their spectra available to
us in electronic form.
Hence, to measure $W_{0}(\CIV )$ 
in these literature objects we were able to apply exactly the same 
procedure as used for the 
galaxies presented in this work. For the objects investigated by Steidel 
et al. (1996a, 1998), 
Lowenthal et al. (1997) and Trager et al. (1997) we measured the 
$W_{0}(\CIV )$ from enlarged tracings. We tested the reliability of
measuring $W_{0}(\CIV )$ from tracings using some 
of our own high-z galaxies. The difference between the two measurement
methods turned out to be $\leq 10 \%$. Although the spectral resolutions of
the different investigations are not exactly the same, the individual
resolutions are sufficiently close to allow a direct comparison
within the accuracy needed here. Table~\ref{tab:mciv_lit} and    
Fig.~\ref{fig:mcivz_fors_lit} 
show that the average \CIV\ values derived from the 
literature high-$z$ spectra are in reasonably good
agreement with the mean values derived for the galaxies investigated in this 
work. However, the scatter of the literature data at high redshift is larger.
Although this larger scatter is presumably dominated by the on average
low S/N of the literature spectra, we cannot exclude that environmental 
effects may influence the evolution of the \CIV\ strength at high redshifts.

The best investigated individual high-$z$ galaxy is, so far, the
gravitationally magnified object MS1512-cb58 ($z= 2.727$) 
(cf. e.g. Yee et al. 1996, Seitz et al. 1998, Pettini et al. 2000, Teplitz et al. 2001, 
Savaglio et al 2002). By measuring the \CIV\ equivalent width on low
resolution spectra and using Eq.~\ref{eq:metalcalib} we obtain 
for this galaxy a metallicity of $0.4 Z_{\odot}$ and $0.2 Z_{\odot}$ 
from P00's and Y96's data, respectively. 
Within our error limits these values are in good agreement with 
the result of Pettini et al. (2000) (who derive 
$0.25 Z_{\odot}$ by comparing the galaxy spectrum with
synthetic starburst galaxy spectra from Leitherer et al. (2001)) and
Teplitz et al. 2001 (who found
$0.32 Z_{\odot}$ from measuring its oxygen abundance
on NIR spectra using the strong line index $R_{23}$ which 
relates (O/H) to the relative abundance of $[$OII$]$, $[$OIII$]$ and 
H$\beta$). This comparison seems to support our assumption that our
calibration of the \CIV\ strength in terms of metallicity is applicable to
high redshift objects at $z=2.7$, although the redshift of MS1512-cb58 is 
too small to estimate the accuracy of the method for 
the interesting $z>3$ objects. 

For four further high-z Lyman-break galaxies Pettini et al. (2001)
determined the oxygen abundance
from NIR spectra using the strong line index $R_{23}$.
Since the relation between $R_{23}$ and O/H has an upper and a lower 
branch these abundances show the well-known two-value ambiguity.
Hence these results do not provide a reliable test of our conclusions.
Nevertheless, in Fig.~\ref{fig:mmet_z_lit} we plot the
allowed ranges of oxygen abundance
for these 4 Lyman-break galaxies as well as for MS 1512-cb58
together with the metallicities of our starburst galaxies (as derived 
from the \CIV\ strength via the calibration described above) as a function
of redshift. This comparison shows that all data are at least mutually 
compatible, 
although for the two highest-z galaxies from Pettini et al. (2001) only the 
lower-branch results are in reasonable agreement with a 
strong increase of metallicity from redshift $\approx 3.2$ to $\approx 2.3$
suggested by our results.

Our results are also in line to those obtained by de Breuck et al. (2000),
who find a qualitative 
increase of metallicity from higher to lower redshift
for a sample of high-z radio galaxies.
Furthermore Pettini et al. (1997) and Savaglio et al. (2000)
report on evidence for a gradual chemical enrichment of the gas producing 
the damped Ly$\alpha $ lines in QSO spectra, although
their trends are only weakly significant.
Compared to Savaglio et al. (2000) we find a zero point 
offset of the metallicity-redshift relation of 
about 0.7 in $\Delta \log Z$ at $z = 2.5$.  Such a difference is not
unexpected since the metal absorbers in damped Ly$\alpha $ systems 
most likely sample the outermost regions of galaxies and therefore 
a different environment than 
the dense interstellar matter of which the massive stars 
seen in starbursts have been formed.

\section{Conclusions}
\label{Conclusions}

Our study shows that the FDF starburst galaxies at 
$z \approx 3.2$ have on average significantly lower \CIV\ equivalent widths 
than starburst galaxies at lower redshifts. In view of the
known close relation of the \CIV\ strength to the metallicity 
in local starburst galaxies, it appears likely that this effect is due to
a significant evolution of the average metallicity in such objects
at high redshifts. 
Using data from well studied local starburst galaxies
we calibrated the \CIV\ strength in terms of the 
heavy element content of these objects. Assuming that this calibration is
applicable to high redshift starburst galaxies we find    
that the mean cosmic metallicity as observed in starburst galaxies has  
increased significantly between the cosmic epochs corresponding
to $z \approx 3.2$ and $\approx 2.3$ 
(when the universe was between 2 and 3 Gyrs old).
If this interpretation of the increase of the \CIV\ absorption
strength  with decreasing redshift is
correct, an intense phase of star formation and evolution 
of massive stars must have occurred during this period. 
At lower redshifts ($z <2.5$) our data indicate little further increase of 
the average metallicity of starburst galaxies. Hence,
the further cosmic chemical enrichment seems to have  been insignificant 
during the last 11 Gyrs. The metallicity evolution indicated by our data 
is in reasonable agreement with published 
theoretical chemical enrichment models due to star formation at 
early cosmological epochs. Our results are 
also in agreement with a metallicity evolution
found in high-z radio galaxies (de Breuck et al. 2000) and the tentative 
evidence
for a gradual chemical enrichment of the gas producing the  
damped Ly$\alpha $ lines (Pettini et al. 1997, Savaglio et al. 2000).

Since the FDF observations provide a fairly complete sample of the
bright starburst galaxies in the observed direction and redshift
ranges, the observed chemical evolution should be characteristic of the
cosmic volumes with the most intense star formation at the corresponding
epochs. However, it is also clear, that our results do not 
apply to all objects and volumes at a certain redshift or epoch. It is,
e.g., well known that the BLR gas of high-redshift QSOs is characterized 
by high metallicities and that no significant 
chemical evolution is observable in these objects up to at least
$z=5$ (see e.g. Hamann and Ferland, 1999, Dietrich et al. 1999,
Dietrich and Wilhelm-Erkens, 2000). Obviously, in the environment
of these early QSOs much star formation and stellar evolution must
have taken place at epochs corresponding to even higher redshifts. It is
also known, however, (and confirmed from the FDF survey) that high-z  
starburst galaxies are much more frequent than bright high-z QSOs.
Hence, the starburst galaxies are expected to be a more
representative tracer of the history of the 
overall cosmic chemical evolution than the QSOs.   

From our data we cannot determine whether at high redshift a 
metallicity-luminosity 
relation does exist, since we do not have any faint objects in our
high-$z$ sample. But it is evident that the high-redshift galaxies
are on average overluminous for their metallicities compared with local 
starburst galaxies. This trend is also found by Pettini et al. (2001) 
and Kobulnicky \& Koo (2000) for Lyman break galaxies. 
From tests on various subsamples we find that the observed dependence of
$W_{0}(\CIV )$ on redshift is not caused by a luminosity effect.
Furthermore we showed that this dependence is also not cause by the possible 
selection effect of preferentially observing galaxies with low dust content and 
hence low metallicity at high redshifts.

Differences in the \SiIV\ to \CIV\ ratios between local and high-$z$
galaxies in our sample suggest differences in the population and star formation
history in the galaxies with $z>3$. Short bursts of star formation may
have been more important (relative to periods of ``continuous star 
formation'') at these early epochs. 

\begin{acknowledgements} 
We are greatly indebted to Drs. M. Pettini and H.K.C. Yee for providing  
the ASCII files of
their high redshift spectra published in Yee et al. (1996) and Pettini et al. 
(1998, 2000). We thank the referee C. Leitherer for valuable comments.
We also want to thank Paranal staff for their support.
This research was supported by the German Science Foundation (DFG)
(Sonderforschungsbereiche 375 and 439).
\end{acknowledgements}

\end{document}